\documentstyle[titlepage,12pt]{article}

\title{Simulations of Solid--on--Solid Models
       of Spreading of Viscous Droplets}

\author{O. Ven\"al\"ainen$^1$, T. Ala--Nissila$^{1,2}$,
 and  K. Kaski$^{1}$ \\
\\ $^1$Tampere University of Technology
\\ Department of Electrical Engineering
\\ P.O. Box 692, FIN--33101 Tampere, Finland \\
\\ $^2$University of Helsinki
\\ Research Institute for Theoretical Physics
\\ P.O. Box 9 (Siltavuorenpenger 20 C)
\\ FIN--00014 University of Helsinki, Finland \\
  {\it and}
\\ Brown University
\\ Department of Physics, Box 1843
\\ Providence, R.I. 02912, U.S.A.
\\
\date{May 26, 1994}
}

\begin{document}

\maketitle

\textheight 22cm
\textwidth 14.5cm

\oddsidemargin 0.96cm
\evensidemargin 0.96cm
\topmargin -0.11cm
\columnsep 0.5in
\raggedbottom
\parindent=0mm

\baselineskip 24pt

\begin{abstract}

We have studied the dynamics of spreading of
viscous non--volatile fluids on
surfaces by Monte Carlo simulations of solid--on--solid (SOS)
models. We have concentrated on the complete
wetting regime, with surface diffusion barriers
neglected for simplicity.  First, we have performed
simulations for the standard SOS model.
Formation of a single precursor layer, and
a density profile with a spherical cap shaped center surrounded
by Gaussian tails
can be reproduced with this model. Dynamical layering, however, only
occurs with a very strongly attractive
van der Waals type of substrate potential.
To more realistically describe the spreading of viscous liquid
droplets, we introduce a modified SOS model. In the new model,
tendency for dynamical layering and the effect of the surface
potential are in part embedded into the dynamics of the model.
This allows a relatively simple description of the
spreading under different conditions, with a
temperature like parameter which strongly influences the
droplet morphologies. Both rounded droplet shapes and dynamical
layering can easily be reproduced with the model.
Furthermore, the precursor width increases proportional to
the square root of time, in accordance with experimental observations.

\end{abstract}

\section{Introduction}

Surface processes such as wetting \cite{deG}, coating
and adhesion depend strongly on phenomena
occuring in the first few layers on top of the substrate.
Recent progress in optical techniques [2-8]
has focussed attention to the
dynamics of spreading of tiny, viscous nonvolatile droplets on
solid surfaces. With ellipsometry, measurements can now be
done on molecular scales. They have revealed
fascinating phenomena previously unaccounted for in macroscopic
droplets, such as dynamical layering and diffusive spreading
of the precursor layer \cite{HesN,HesP2,Fra,DeC}.
\medskip

In the experiments, the most common liquids have been tetrakis
(2--ethyl\-hex\-oxy)--silane, squalane, and polydimethylsiloxane
(PDMS).  It has been shown \cite{HesN}
that the thickness profiles of tetrakis and
PDMS  droplets on a silicon wafer exhibit
strikingly different shapes under spreading. On a
``low energy'' surface (with a relatively low solid--vacuum
surface tension \cite{deG})
tetrakis produces clear
dynamical layering (Fig. 1(a)), i.e. stepped shapes of exactly
one molecular layer in thickness. In contrast,
on the same substrate the spreading of PDMS proceeds
by a fast evolving precursor of about
one molecular layer in thickness (Fig. 1(b)).
%
%
\medskip

On the other hand, the same liquid on {\it different
surfaces} has been shown to form distinct
morphologies \cite{HesP2}. On a ``high energy'' surface
(with a relatively high surface tension and adsorption energy
\cite{deG})
PDMS developes dynamical layering, too, in analogy to tetrakis.
On a low energy surface the formation of a single precursor layer
with a central cap is observed \cite{HesP2}. Other experiments show
\cite{Val,DeC} that PDMS droplets form stepped,
dynamically layered profiles on two different
silicon wafers (Fig. 1(c)). In addition, a recent experiment
of PDMS spreading on a silver substrate shows \cite{Alb}
the formation of a profile with spherical cap shaped
 center and Gaussian tails at late times (Fig. 1(d)).
\medskip

On the theoretical side progress has been more moderate. The
continuum hydrodynamic approach \cite{deG,Leg,Tan,Joa}, while
being able to qualitatively explain the precursor film
and its eventual diffusive ($\sim t^{1/2}$) growth \cite{Joa}
is simply not valid for the measurements
on scale of {\AA}ngstr\"oms.  The first theoretical model
to successfully account for dynamical layering
was given by Abraham {\it et al.} \cite{Abr1}.
They used the two dimensional horizontal
solid--on--solid (SOS) model, which describes
coarse--grained layers of fluid emanating from a droplet ``reservoir''.
Dynamical layering was observed to occur
with an attractive van der Waals type of surface potential
in the complete wetting regime \cite{Abr1,Abr2}.
The model also predicts diffusive behaviour for the bulk layers of
the droplet on a high energy surface, but {\it linear} time
dependence for the precursor layer.
\medskip

A different approach was recently taken by
de Gennes and Cazabat \cite{deG2}, who proposed an analytic model
of an incompressible, stratified droplet.
The stepped morphology of the spreading droplet is
postulated {\it a priori}, and spreading is
assumed to proceed by a longitudinal flow of the molecular layers, and
a transverse permeation flow between them.
The model introduces an effective Hamaker constant, which has been
fitted \cite{Val} to experimental data on stepped PDMS droplets.
\medskip

On a more microscopic level, molecular dynamics simulations
have been performed \cite{Hau,Yan,Nie1}.
Yang {\it et al.} \cite{Yan} studied droplets of Lennard--Jones (LJ)
particles and dimers on corrugated surfaces,
surrounded by a vapor phase. They observed terraced droplet
shapes in the complete wetting regime, but obtained logarithmically
slow growth for the layers. In contrast, Nieminen {\it et al.}
\cite{Nie1} were able to qualitatively observe a crossover
from ``adiabatic'' to ``diffusive'' spreading for LJ droplets
on a smooth surface. To account for the chainlike molecular
structure
of some of the liquids, Nieminen and Ala--Nissila \cite{Nie2}
recently carried out a systematic study of droplets of
LJ solvent particles and flexible oligomers on smooth surfaces.
By varying the strength of the surface potential, they were able
to qualitatively conclude that
the actual molecular structure of the liquid can play
an important role in the formation
of the density profiles, even beneath the entanglement
regime \cite{deGB}. Namely, relatively weaker substrate
interactions and
longer chains favour more rounded droplet shapes, due to the
mixing of the molecular layers under spreading. On the other
hand, a very strong surface attraction promotes layer separation and
leads towards stepped shapes. In the latter case, the spreading process
qualitatively proceeds in analogy to the ideas of Ref. \cite{deG2},
with layers of fluid ``leaking'' at the edges of the steps.
They also observed a crossover from
``adiabatic'' to ``diffusive'' behaviour of the precursor film and
developed a scaling form for its time dependence.
These results are in good qualitative agreement with experiments;
however, droplets consisting of just a few thousand molecules
were simulated, while experimental droplets
are microscopic in the vertical direction {\it only}.
\medskip

In the present work we have taken to approach the problem of
droplet spreading on a more coarse--grained level, in order
to further clarify the role of the effective viscosity
of the liquid, and the surface potential in the droplet
morphologies. We have carried out systematic Monte Carlo
simulations with two different SOS models to study
the dynamics of the spreading process.
In these models, the droplet is assumed to consist of
effective, coarse--grained blocks of molecules \cite{Abr1}.
The size of these blocks can be highly anisotropic between
the vertical and
horizontal directions, depending on the details of the interactions.
All the simulations have been done in the complete wetting
regime in the canonical ensemble, where the final state of the
droplet is a surface monolayer.
No evaporation from the surface is allowed, and in this work
the surface diffusion barriers have been neglected for simplicity.
\medskip

The first model that we have studied is the standard vertical SOS
model \cite{Lea,Hei}. Besides computing density profiles
of the droplets, we have also followed the time evolution
of the width of the precursor film $r(t)$, and the droplet
height. Our results demonstrate that
with a strongly attractive short range substrate potential, only
a single precursor layer evolves, with the center of the droplet
assuming a spherical cap shape rather than a Gaussian profile.
At later times, a surface monolayer forms. These results are analogous
to the observations in Ref. \cite{Hei}.
In addition, formation of a spherical cap with
Gaussian tails occurs with relatively weak attractive short range
substrate potentials.  To obtain stratified droplet shapes
corresponding to dynamical layering, however, an extremely strong
long range van der Waals type of surface potential must be
used. In addition, the behaviour of the precursor film was
not observed to follow the correct dynamical behaviour.
This is due to the somewhat unrealistic dynamics of the model.
\medskip

To better take into account the
fluid--like nature of the spreading droplets,
we have developed a modified
SOS model in which the tendency for dynamical layering
and the effect of the surface potential have been partially embedded
into the {\it dynamics} of the model, in analogy with
some of the ideas presented by de Gennes and Cazabat \cite{deG2}.
In its simplest form, the model enables a
description of the spreading process with a single temperature
like parameter. When varied, this parameter strongly influences
the droplet morphologies and reproduces both rounded and stepped
droplet shapes. In particular, dynamical layering can be
reproduced without explicitly including a long range
van der Waals substrate attraction.
When such an attraction is added the range of layering
is further enhanced, as expected.
Our quantitative results also show that the precursor width
increases proportional to the square root of time.
A preliminary account of these results has been
published in Ref. \cite{Ven1}.

\section{Standard Solid--on--Solid Model of Spreading Dynamics}

\subsection{Definition of the Model}

The standard solid--on--solid (SOS)
model used in our simulations is defined by the following
Hamiltonian \cite{Ven2}:

$$H (\lbrace h \rbrace)={{\tau }\over 2}\sum_{i,j}\min (h_i,h_j) +\tau
 \sum_{i=2}^{N} h_i+ \sum_i V(h_i) ,\eqno (1)$$

where in the first term $\tau < 0$ is the nearest neighbour
interaction parameter and the summation goes
over the four nearest neighbours of the column $i$
whose height is $h_i$. The heights, which
measure the average concentration of particles and obey
a vertical SOS restriction \cite{Lea} are integers.
The second term on the right hand side of Eq. (1) describes the vertical
interaction between the effective molecules.
$V(h_i)$ denotes the strength of an attractive, height dependent
substrate potential at site $i$. $V(h_i)$ is defined as

$$V(h_i)={A\over {h_i^3}} + B h_i \delta_{h_i,1} ,
\eqno (2)$$

where $A<0$ is an effective Hamaker constant, and $B<0$ describes the
interaction between the substrate and the fluid.
\medskip

In the model, the effective particles follow standard Monte Carlo
(MC) Metropolis dynamics with transition probability

$$P(h_i\to h_i')=\min \lbrace 1, e^{-\Delta H/k_BT}\rbrace ,\eqno (3)$$

where $\Delta H $ is the energy difference
between the final ($h_i'$) and
initial ($h_i$) states (with the SOS restriction obeyed).
Time is measured in units of MC steps per site (MCS/s).
The initial configuration of the droplet is usually chosen to
be a three--dimensional cube, which after a transient time
assumes its characteristic shape. During the spreading
stage, we calculate the density profile of the droplet
(in analogy to the experiments), the time dependence of the
precursor width $r(t)$ and the height of the droplet.
All the simulations have been done without allowing evaporation
from the surface; thus the final state is a surface monolayer.
In this work the activation energy for surface diffusion
of isolated particles is neglected, and thus the model may
not be applicable to some high energy surfaces.

\subsection{Results for the Standard Model}

First, we study the spreading dynamics
without the van der Waals potential by setting $A=0$ in Eq. (2).
In this case, the results are not very sensitive to the droplet
size, and in most cases a $11\times 11\times 10$
droplet was used, where ten is the initial height.
We also fix $\tau =-0.05$ in this section.  With a relatively
strong surface attraction, only a
single precursor film is formed, with a rounded central cap.
At the edge of the film the migration
of the effective molecules causes the formation of a monolayer
of separate particles.
This happens typically in the range $\vert B/\tau \vert =2.0-4.0$,
and $\vert k_BT/\tau \vert =0.6 - 1.0$. These results are in
complete agreement with the qualitative observations of Ref.
\cite{Hei}. Additionally, we find that the time development of the
precursor film approximately follows $r(t)\sim t^{0.14}$, in contrast
to $t^{1/2}$. If we then lower the surface attraction
and set $B=\tau $, and $\vert k_BT/\tau \vert =0.6$,
the precursor film vanishes and the droplets
assume a rounded, spherical shape with Gaussian tails as shown in Fig. 2.
\medskip

Next, we study the influence of the long range van der Waals type
potential. For a moderately large $A$, the results are virtually
indistinguishable from those above. Namely, for $A=B=-0.15$, a single
precursor film forms while for $A=B=\tau$ the film disappears
($\vert k_BT/\tau \vert = 0.6$). However, for a very large surface
attraction $A=B=-1.0$, the droplets show indications of separation
of layers. At $\vert k_BT/\tau \vert=1.0$ shown in Fig. 3(a),
a small central cap is formed at submonolayer heights. When
temperature is further lowered to 0.6, three separate layers
can be observed (Fig. 3(b)). We have calculated
$r(t)\sim t^\alpha $ for the lower temperature case, and
find $\alpha \approx 0.17$ for both $A=B=-0.15$ and $-1.0$.
We also find that at very late times, the height of the
submonolayer film decreases approximately as $t^{-2}$. This describes
the decrease of the almost two dimensional island,
which leaks into a surface monolayer. This monolayer which spreads
diffusively consists of all
the particles, which are two or more lattice sites apart from the
cluster.
 Thus the precursor film width decreases and the areal density of the
 monolayer correspondingly increases.
We note that due to the lack of surface diffusion activation
energy, the case of larger surface attraction enhances the rate
of spreading, which is opposite to what happens on high energy
surfaces \cite{HesP2,Nie2}.
\medskip

{}From the results presented above, it is clear that the standard
SOS model can produce dynamical layering only for exteremely
large values of the long range surface attraction in the model.
In fact, although
standard SOS models have been very useful in describing solid
surfaces and their roughening \cite{Lea}, and even growth of surface layers
\cite{Ven2}, it is clear that they are less
realistic for fluid spreading. For example, the diffusion length of the
effective particles  is not well--defined, due to the fact that
very large height differences between nearest neighbour are
allowed. Second, nonmonotonic variations in the density profiles
are possible, although not significant for the parameter range
used in the present work. Due to these limitations, in the next
section we shall introduce a modified SOS model which better
accounts for the fluid--like nature of the spreading droplets.

\section{Modified Solid--on--Solid Model of Spreading Dynamics}

\subsection{Definition of the Model}

The basic idea behind the modified SOS model is to more realistically
describe the viscous flow of the droplets. We have done this by
modifying the {\it dynamics} of the standard SOS model defined
by Eqs. (1)--(2) in the following way.
First, the effective molecules are not allowed to climb over
height barriers, or jump down over terrace edges. For example,
two individual particles located initially at different
sites on the surface can never form a column of height $h_i=2$.
Second, when the difference
between the neighbouring initial and final sites is $(h_i-h_i')\ge 2$,
a {\it transient} non--SOS excitation (a hole or vacancy)
can be created into the initial column by a particle which jumps
to the neighbouring position with a probability

$$P=\min \lbrace
\Theta (h_i-h_j-1/2),e^{-{\Delta H}_{t}/k_BT}\rbrace \eqno (4),$$

where $\Theta (x)$ is the Heavyside step function.
The energy difference $\Delta H_t$ is
calculated for this {\it transient} state. The hole left behind is
immediately filled by the column of particles above it, which are all
lowered by a unit step. This means that $\Delta H_{t}>\Delta H$
(as calculated for the final state). We note that the
effective dynamics used here does not satisfy microscopic reversibility
at all times.
\medskip

The motivation for the unusual dynamics comes from the physics of
dissipative
spreading of viscous fluids on attractive surfaces. Microscopic
calculations have revealed \cite{Nie2} that at least in the
layered state, the droplets spread by the viscous flow of
rather well separated layers, which ``leak'' on the surface
at the terrace edges.
Furthermore, the analytic model for stratified droplets by
de Gennes and Cazabat \cite{deG2} assumes, that the horizontal
layers of fluid flow with friction, and that there is a narrow region
of permeation flow at the terrace edges.
In our model, this is roughly described by the creation of the
transient hole at the step edges. Since this is temperature
controlled (cf. Eq. (4)), we expect that the model be able to
describe droplet spreading under a variety of different conditions.

\subsection{Results}

First, we present results for $A=B=\tau=0$, which corresponds
to an infinite temperature. The effective dynamics in the
model renders even this simplest case nontrivial.
As seen in Fig. 4, at early and intermediate stages of
spreading the droplet assumes a non--Gaussian, spherical shape with
Gaussian tails surrounding the center
of the droplet. At early times the spreading is relatively
fast, but slows down at later times as the droplet
density profile approaches a Gaussian shape at
submonolayer heights. In contrast, we note that
an infinite temperature for the standard SOS model always leads
to Gaussian shaped profiles. At late times, the width of
the precursor film approaches the expected $t^{1/2}$ behaviour.
\medskip

Next, we set  $A=B=0$, $\tau =-0.05$ and let the
temperature ratio $ k_BT/\tau  $ vary.
For the lowest temperature of 0.6, {\it dynamical layering}
of the density profiles is immediately evident in Fig. 5(a).
For larger droplets, up to five or six layers can easily
be detected (Fig. 5(b)). In this case the droplet maintains a rather
compact shape and the surface
monolayer formation is relatively slow.
At later stages (not shown in Fig. 5) a flat
profile is formed with a tiny central cap at the center.
This cap disappears at approximately after 6000 MCS/s.
The precursor width follows $t^{0.50(3)}$. We also
studied the spreading of an initially ridge--shaped
droplet of size $120\times 11\times 10$, where the
precursor width follows $t^{0.75}$ rather than $t^{1/2}$. In
this case the spreading is highly anisotropic
and rather different from the three--dimensional droplet.
\medskip

When temperature is raised, the layer formation weakens and
droplets become more spherical. For $\vert k_BT/\tau \vert =0.8$
(Fig. 6) and 1.0, the  precursor film grows as $t^{0.49(7)}$ and
$t^{0.47(6)}$, respectively. The surface
monolayer formation due to enhanced
thermal fluctuations is faster, and for the higher temperature
the whole precursor film tends to break down.
In Fig. 7 we show the time development of the droplet height at the
three different temperatures. The fast spreading rate, and consequently
the enhanced edge flow leading to more rounded shapes
at higher temperatures is clearly reflected in the results.
\medskip

Next we study the spreading with a long range van der Waals type of surface
potential with $\vert A/B \vert =1$.
 In this case simulations have been performed
 for $B=-0.025$, $B=-0.05=\tau$,
$B=-0.07$, and  $B=-0.2$.
The temperature was set to $\vert k_BT/\tau \vert=0.6$.
In our model with modified dynamics, the inclusion of the
van der Waals energy enhances the tendency for dynamical layering,
as demonstrated in Fig. 8.
Furthermore, the results indicate clearly that
with $B=-0.2$ the precursor films breaks down, giving
for the time development of the precursor film width the
behaviour $\sim t^{0.3}$. With the other choices, however, the time dependence
of the precursor width follows $t^{1/2}$ rather accurately.
For $B=-0.2$ spreading becomes
unrealistically fast due to the lack of the surface diffusion
activation barriers, which have been set to zero.
As mentioned earlier, we do not expect the model to be applicable
to high energy surfaces if the diffusion barriers control
the dynamics.

\section{Discussion of the Results and Comparison with Experiments}

Next, we analyze the density profiles and make comparisons
with relevant experiments [2-8],
whenever possible. Due to the coarse--grained nature of the
SOS models quantitative comparisons may not be meaningful,
however. The aim here in part has been to eludicate the
physical processes behind different droplet morphologies, and
thus we will also discuss our results in connection to
other coarse--grained models \cite{Abr1,Abr2,deG2,Hei},
and microscopic simulations \cite{Yan,Nie1,Nie2}.

\medskip
First, we discuss results for the standard SOS model. The model
qualitatively reproduces single precursor film formation with
a relatively strong attractive short range substrate potential,
while with a weaker potential, rounded droplet shapes
with Gaussian tails appear. This is in accord with
the notion that increased surface attraction should
enhance the precursor film \cite{HesP2,Alb,Abr1,Abr2}.
The precursor film was also qualitatively observed in
the simulations of Ref. \cite{Hei}. Also, with the inclusion
of a long range van der Waals potential, the droplets
exhibited tendency towards dynamical layering. This is
in accordance with the horizontal SOS model results
of Refs. \cite{Abr1,Abr2}, and qualitatively with the
microscopic picture as well \cite{Nie1,Nie2}.
Although the standard SOS model
can reproduce density profiles resembling experimentally
observed ones (Fig. 3(a) \cite{HesP1} and Fig. 3(b) \cite{HesN,Val})
the strength of the long range potential must be considered
unrealistically large as
compared with the particle--particle interactions on the
surface level. Most importantly, the time dependence
for the width of the precursor film does not follow the
expected $t^{1/2}$ behavior, which reflects the unrealistic
nature of the dynamics in the model.
\medskip

In contrast, the modified SOS model yields results which seem
much more realistic. At low temperatures, stepped droplet shapes
are obtained (Figs. 5(a) and (b)). On the other hand, when
particle motion from one layer to another is more likely,
rounded droplet shapes occur (Figs. 4 and 6). This is
in qualitative accord with conclusions from coarse--grained
models \cite{Abr1,Abr2,deG2}
and microscopic calculations \cite{Nie1,Nie2}.
Furthermore, the inclusion of a van der Waals potential
further enhances the range of layering, as expected. The results
are also in qualitative agreement with experiments \cite{HesP2,Alb,Val}
where increasing the viscosity of the liquid enhances
layering, and layering is also more prominent on high
energy surfaces. Most importantly, the model reproduces
the expected $t^{1/2}$ dependence for the width of the
precursor layer. However, it should be noted again that since
the model neglects diffusion barriers, our results may
not be applicable for all cases.
\medskip

The remarkable feature of the modified SOS model is that by just
varying a single temperature like parameter, droplet morphologies
from rounded to stepped shapes can be easily reproduced.
This can be understood based on the physical analogy behind
the modified dynamics \cite{deG2}. High effective temperatures
in the model correspond to enhanced flow at the edges, and
reduced viscocity and surface interactions allowing
transient hole formation with relative ease. On the other
hand, lowering the temperature suppresses this hole
formation which is analogous to suppressing interlayer
flow in the phenomenological model of de Gennes and
Cazabat \cite{deG2}. In fact, our model gives further support
to the physical ideas behind their model regarding
dynamical layering.

\section{Summary and Conclusions}

To summarize, in this work we have tried to unravel physical
processes behind  different morphologies observed in the
spreading of tiny liquid
droplets on surfaces. The present approach
of using coarse--grained descriptions of the fluid is
complementary to other approaches, and can indeed
give a lot of physical insight to the problem. In particular,
by introducing a new, more realistic coarse--grained lattice
SOS model of fluid spreading we have demonstrated how the
interplay between the
internal viscosity of the fluid and the nature of the
surface attraction can produce a number of
different droplet shapes, in qualitative agreement with
experiments. Our results also give further support to
ideas behind the analytic models, in particular that of
Ref. \cite{deG2} regarding the spreading of terraced,
stratified droplets. They are also in agreement with
microscopic studies of Refs. \cite{Nie1,Nie2}. However,
the experimental situation remains complicated and should
not be oversimplified -- there exist very few attempts to
systematically study the effects of surface attraction,
microscopic structure of the liquid, finite size
effects and other factors, which can play an important
role in spreading dynamics. We hope that the present work
inspires more systematic studies in these directions.
\medskip

Acknowledgements: We wish to thank R. Dickman, S. Herminghaus,
and J. A. Nieminen for useful discussions,
and the Academy of Finland for financial support. We also wish to thank
P. Leiderer for sending his data on the density profiles of PDMS spreading
on silver.

\pagebreak
\Large
{\bf Figure captions}

\normalsize
\vskip1cm
\baselineskip 18pt

Fig. 1.
(a) Experimental thickness profiles of tetrakis on a ``low
energy'' silicon surface \cite{HesN}. Dynamical layering occurs at late
times.
(b) Thickness profiles of PDMS on the same surface
(the ``Mexican hat'' shape) \cite{HesN}. For even smaller droplets,
the central cap becomes more distinct
and a Gaussian shape is assumed at latest times \cite{HesP1}.
(c) Thickness profiles of PDMS on a silicon wafer covered by a
grafted layer of trimethyl groups. This surface behaves as
a ``low energy'' surface \cite{DeC}, with strong dynamical layering
occuring.
(d) Thickness profiles of PDMS on a silver surface after 50, 80, 150, and
360 minutes \cite{Alb}. The submonolayer profiles can be fitted by
a spherical cap, with Gaussian convolution at the edges.
\vskip1cm

Fig. 2.
Density profile for the standard SOS model, with $A=0$, $B=\tau$, and
$\vert k_BT/\tau \vert = 0.6$. The droplet assumes a rounded profile,
with no precursor film (cf. Fig. 1(d)). The initial height of the droplet
is 41 units, and the result is an average
over 50 runs after 25 000 MCS/s. The solid line
indicates a Gaussian fit.
\vskip1cm

Fig. 3.
Spreading droplet profiles for the standard SOS model with $A=B=-1.0$,
and (a) $\vert k_BT/\tau \vert=1$. At submonolayer heights,
a central cap forms.
The curves correspond to 800, 1200, 1400 and 1600 MCS/s,
and are averages over 500 runs. The
initial size of the droplet is $11\times 11\times 10$.
(b) Results for $\vert k_BT/\tau \vert=0.6$. Dynamical layering
tends to occur, with three layers visible.
The curves correspond to  1600,
2200, and 3800 MCS/s, with averages over 300--400 runs.
\vskip 1cm

Fig. 4.
The density profile of a spreading droplet for the modified
SOS model at an infinite temperature, after 650 MCS/s.
A distinctly non--Gaussian, spherical shape occurs above
monolayer heights, as evidenced by a Gaussian fit shown with
a solid line.  The initial height of the droplet is 41 units, and
the profile is an average over 500 runs.
\vskip1cm

Fig. 5.
Density profiles for the modified SOS model at $\vert k_BT/\tau
\vert=0.6$. {\it Dynamical layering} of the profiles
occurs after an initial transient. (a) Results for a
$11 \times 11 \times 10$ droplet, with curves corresponding
to 1800, 2800, and 3800 MCS/s. (b)
Results for a $11 \times 11 \times 40$ droplet, with curves
at 5800, 6800, and 8000 MCS/s.
All profiles are averages over 500 runs.
\vskip1cm

Fig. 6.
Density profiles for the modified SOS model at $\vert k_BT/\tau \vert =0.8$.
A shoulder and a central cap are clearly seen, with a spherical
cap shaped center and a Gaussian foot at late times.
The curves correspond to 2000, 2200, and 2600 MCS/s.
Results are averages over 500 runs.
\vskip1cm

Fig. 7.
The height of the droplets for the modified SOS model
at the three different temperatures studied, as a
function of time. Filled squares correspond to
$\vert k_BT/\tau \vert =0.6$,
open circles to 0.8, and filled circles to 1.0. Breaks
in the curves correspond to separate layers. While the spreading
rate strongly increases as a function of temperature, the
dynamical layering considerably weakens and droplets become
more rounded.
\vskip1cm

Fig. 8.
Density profiles for the modified SOS model, with $\vert A/B \vert=1$,
$\vert k_BT/\tau \vert=0.6$, and $B=-0.2$. A stronger
van der Waals potential clearly enhances dynamical layering
(cf. Fig. 5). See text for details.
\vskip1cm

\pagebreak

\end{document}